\documentstyle[psfig]{mn}

\newif\ifAMStwofonts

\ifoldfss
\ifCUPmtlplainloaded \else
\NewTextAlphabet{textbfit} {cmbxti10} {}
\NewTextAlphabet{textbfss} {cmssbx10} {}
\NewMathAlphabet{mathbfit} {cmbxti10} {} 
\NewMathAlphabet{mathbfss} {cmssbx10} {} 
\fi
\ifAMStwofonts
\ifCUPmtlplainloaded \else
\NewSymbolFont{upmath} {eurm10}
\NewSymbolFont{AMSa} {msam10}
\NewSymbolFont{AMSa} {msam10}
\NewMathSymbol{\upi}     {0}{upmath}{19}
\NewMathSymbol{\umu}     {0}{upmath}{16}
\NewMathSymbol{\upartial}{0}{upmath}{40}
\NewMathSymbol{\leqslant}{3}{AMSa}{36}
\NewMathSymbol{\geqslant}{3}{AMSa}{3E}

 \let\le=\leqslant
 
\fi
\fi
\fi 

\ifnfssone
\newmathalphabet{\mathit}
\addtoversion{normal}{\mathit}{cmr}{m}{it}
\addtoversion{bold}{\mathit}{cmr}{bx}{it}
\newmathalphabet{\mathbfit} 
\addtoversion{normal}{\mathbfit}{cmr}{bx}{it}
\addtoversion{bold}{\mathbfit}{cmr}{bx}{it}
\newmathalphabet{\mathbfss} 
\addtoversion{normal}{\mathbfss}{cmss}{bx}{n}
\addtoversion{bold}{\mathbfss}{cmss}{bx}{n}
\ifAMStwofonts
\ifCUPmtlplainloaded \else
\UseAMStwoboldmath
\makeatletter
\new@mathgroup\upmath@group
\define@mathgroup\mv@normal\upmath@group{eur}{m}{n}
\define@mathgroup\mv@bold\upmath@group{eur}{b}{n}
\edef\UPM{\hexnumber\upmath@group}
\new@mathgroup\amsa@group
\define@mathgroup\mv@normal\amsa@group{msa}{m}{n}
\define@mathgroup\mv@bold\amsa@group{msa}{m}{n}
\edef\AMSa{\hexnumber\amsa@group}
\makeatother
\mathchardef\upi="0\UPM19
\mathchardef\umu="0\UPM16
\mathchardef\upartial="0\UPM40
\mathchardef\leqslant="3\AMSa36
\mathchardef\geqslant="3\AMSa3E

 \let\le=\leqslant

\fi
\fi
\fi 

\ifnfsstwo
\DeclareMathAlphabet{\mathbfit}{OT1}{cmr}{bx}{it}
\SetMathAlphabet\mathbfit{bold}{OT1}{cmr}{bx}{it}
\DeclareMathAlphabet{\mathbfss}{OT1}{cmss}{bx}{n}
\SetMathAlphabet\mathbfss{bold}{OT1}{cmss}{bx}{n}
\ifAMStwofonts
\ifCUPmtlplainloaded \else
\DeclareSymbolFont{UPM}{U}{eur}{m}{n}
\SetSymbolFont{UPM}{bold}{U}{eur}{b}{n}
\DeclareSymbolFont{AMSa}{U}{msa}{m}{n}
\DeclareMathSymbol{\upi}{0}{UPM}{"19}
\DeclareMathSymbol{\umu}{0}{UPM}{"16}
\DeclareMathSymbol{\upartial}{0}{UPM}{"40}
\DeclareMathSymbol{\leqslant}{3}{AMSa}{"36}
\DeclareMathSymbol{\geqslant}{3}{AMSa}{"3E}

 \let\le=\leqslant

\fi
\fi
\fi 

\ifCUPmtlplainloaded \else
\ifAMStwofonts \else 
\def\upi{\pi}
\def\umu{\mu}
\def\upartial{\partial}
\fi
\fi

\title{A comprehensive CCD photometric study of the open cluster NGC 2421}
\author[R. K. S. Yadav and Ram Sagar]
{R. K. S. Yadav$^{1}$\thanks{E-mail: rkant@iucaa.ernet.in} and Ram Sagar$^{2}$\thanks{E-mail: sagar@upso.ernet.in}\\
$^{1}$Inter-University Centre for Astronomy and Astrophysics, Ganeshkhind,
Pune 411 007, India\\
$^{2}$State Observatory, Manora Peak Nainital 263 129, India}
\date{Accepted ---------.
Received ---------;
}
\pagerange{\pageref{firstpage}--\pageref{lastpage}}
\pubyear{2004}
\begin{document}
\maketitle
\label{firstpage}
\begin{abstract}
              We present the $UBVRI$ CCD photometry in the region of the open cluster NGC 2421.
              Radius of the cluster is determined as $\sim$ 3$^\prime$.0 using stellar 
	      density profile. Our Study indicates that metallicity of the cluster is $Z \sim$ 
	      0.004. The reddening $E(B-V) = 0.42\pm$0.05 mag is determined using two colour 
	      $(U-B)$ versus $(B-V)$ diagram. By combining the 2MASS $JHK$ data with the 
	      optical data we determined $E(J-K) = 0.20\pm$0.20 mag and $E(V-K) = 1.15\pm$0.20 
	      mag for this cluster. Colour-excess diagrams show normal interstellar 
	      extinction law in the direction of the cluster. We determined the distance of the 
	      cluster as 2.2$\pm$0.2 Kpc by comparing the ZAMS with the intrinsic CM diagram of 
	      the cluster. The age of the cluster has been estimated as 80$\pm$20 Myr using 
	      the stellar isochrones of metallicity $Z = 0.004$. The mass function slope 
	      $x = 1.2\pm0.3$ has been derived by applying the corrections of field stars 
	      contamination and data incompleteness. 
	      Our analysis indicate that the cluster NGC 2421 is dynamically relaxed. 
\end{abstract}


\section{Introduction}

 The investigation of young open star clusters provides us a powerful tool to 
 understand the structure and history of star formation in our Galaxy. In order 
 to fully exploit the information provided by open clusters we must know their 
 accurate ages, distances, reddenings, metal abundances and stellar contents. 
 For this, multicolour CCD photometric observations have proved to be very useful. 
 With the development 
 of more accurate stellar models it has been possible to provide a better estimate 
 of the cluster ages simply by comparing theoretical isochrones with the observed 
 CCD colour-magnitude (CM) diagrams. So, In this paper we have considered an open 
 cluster NGC 2421 with 
 the aim of presenting new accurate CCD photometry. From this photometry we 
 select photometric members and derive several fundamental parameters, such as 
 distance, interstellar reddening, metallicity and age as well as luminosity and 
 mass function.
 
 The young open cluster NGC 2421 = C0734$-$205 ($\alpha_{2000} = 07^{h}36^{m}16^{s}$, 
 $\delta_{2000}=-20^{d}36^{\prime}44^{\prime\prime}$; $l = 236^{\circ}.24$, 
 $b = 0^{\circ}.08)$ is classified as a 
 Trumpler class I2m by Ruprecht (1966). This cluster was first studied by Moffat 
 \& Vogt (1975) photoelectrically and derived a distance of about 1.87 Kpc having 
 $E(B-V)=$ 0.47$\pm$0.05 mag and age less than 10$^{7}$ years. Ramsay \& Pollacco 
 (1992) also studied this cluster using CCD photometry and found a colour excess 
 $E(B-V)=$ 0.49$\pm$0.03 mag but a distance of 2.75 Kpc. To our 
 knowledge no other studies have been carried out for the cluster NGC 2421 so far.

 The layout of the paper is as follows. In Sec. 2 we briefly describe the observations 
 and data reduction strategies as well as comparison with the previous photometry. Sec. 3 
 is devoted on the detailed analysis of the present photometric data for the determination 
 of cluster parameter. Finally, Sec. 4 summarizes the main results of the paper.

\begin{figure*}
\centering
\hspace{0.7cm}\psfig{figure=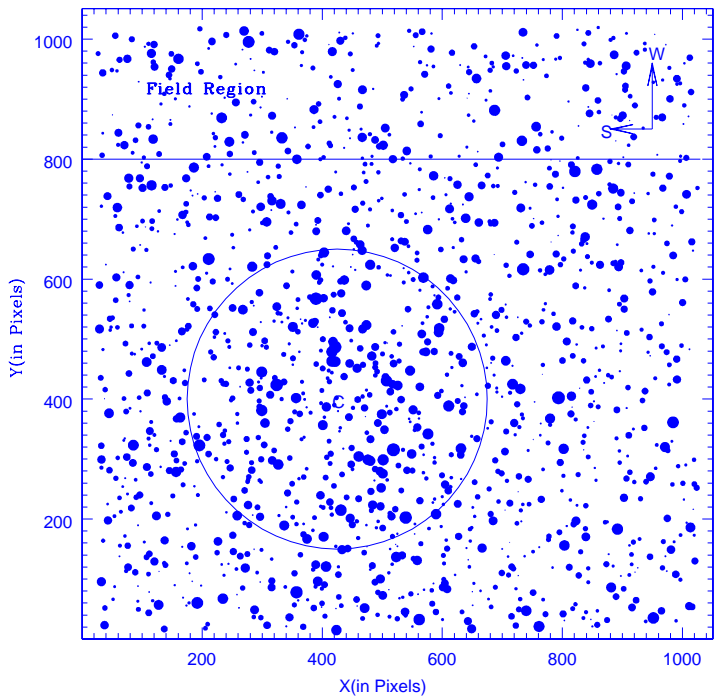,width=10cm,height=10cm}
\caption{Finding chart of the stars in the cluster NGC 2421. The (X, Y) coordinates are 
in pixel units corresponding to 0$^{\prime\prime}$.72 on the sky. Direction is indicated 
in the map. Filled circles of different sizes represent brightness of the 
stars. Smallest size denotes stars of $V$$\sim$20 mag. Open circle having centre at 'C' 
in the chart represent the cluster size.}
\end{figure*}
   
\section{Observations and data reduction}

 $UBVR_cI_c$ CCD photometry was performed for the cluster NGC 2421 on 24/25 February 
 2003 at State Observatory Naini Tal with the 104-cm telescope (f/13) and a 2K$\times$2K CCD 
 (24-$\mu$m pixels). The scale was 0$^{\prime\prime}$.36 pixel$^{-1}$, giving 12.$^{\prime}$6 
 on a side. The read out noise for the system was 5.3 e$^{-}$, while the gain was 
 10 e$^{-}$/ADU. Table 1 lists the log of our CCD observations. To improve the S/N ratio, 
 observations were made in 2$\times$2 pixel binning mode while 2 to 3 deep exposures 
 were taken for the accurate photometric measurements of faint stars. Many bias frames 
 were taken during 
 the observations for removing the bias level in the images. Flat-field exposures were made 
 of the twilight sky in each filter. Fig. 1 shows the finding chart for 
 the stars brighter than $V = 20$ mag in the cluster NGC 2421. We observed the standard 
 area PG 1633$+$099 (Landolt 1992) several times during the night for the purpose of 
 determination of atmospheric extinction coefficients and photometric calibration of the 
 CCD system. 
 The brightness and colour range of the standard stars are 13$\le V \le$15 and 
 $-$0.2$\le (V-I) \le$1.1 respectively. So, the standard stars in this area provide a good 
 magnitude and colour range, essential to obtain reliable photometric transformation.

   The data were reduced using the computing facilities available at Inter-University Centre 
   for Astronomy and Astrophysics (IUCAA) Pune, India. Corrections to the raw data for bias and 
   flat-fielding were performed  
   using the standard IRAF routines. The CCD frames of the same exposure for a given filter 
   were combined to improve the statistics of the faintest stars. Stellar magnitudes were 
   obtained by using the DAOPHOT software (Stetson 1987, 1992)
   and conversion of the raw instrumental magnitudes into those of standard photometric 
   system were done using procedures outlined by Stetson (1992). The instrumental magnitudes 
   were derived through Point Spread Function (PSF) fitting using DAOPHOT. To determine the 
   PSF, we used several well isolated stars for the entire frame. Several stars brighter than 
   $V = 11.0$ mag could not be measured as they saturated even on the shortest frame.

   For translating the instrumental magnitude to the standard magnitude, the calibration 
   equations derived using least square linear regression are as follows:

\begin{center}
   $u=U+4.95\pm0.01-(0.02\pm0.01)(U-B)+0.62X$

   $b=B+3.41\pm0.01-(0.05\pm0.01)(B-V)+0.28X$

   $v=V+3.05\pm0.01-(0.09\pm0.01)(B-V)+0.17X$

   $r=R+2.96\pm0.01-(0.02\pm0.01)(V-R)+0.12X$

   $i=I+3.29\pm0.01-(0.07\pm0.01)(R-I)+0.10X$
\end{center}

   where $U, B, V, R$ and $I$ are the standard magnitudes and $u, b, v, r$ and $i$ are
   the instrumental aperture magnitudes
   normalised for 1 second of exposure time and $X$ is the airmass. We have ignored the
   second order colour correction terms as they are generally small in comparison to
   other errors present in the photometric data reduction. The errors in zero points and 
   colour coefficients are $\sim$ 0.01 mag. The errors in magnitude and colour are 
   plotted against $V$ magnitude in Fig. 2 and the mean values of the errors are listed 
   in Table 2. The final photometric data are available in electronic form at the WEBDA 
   site \footnote{\it http://obswww.unige.ch/webda/} and also from the authors.
\begin{figure}
\centering
\psfig{figure=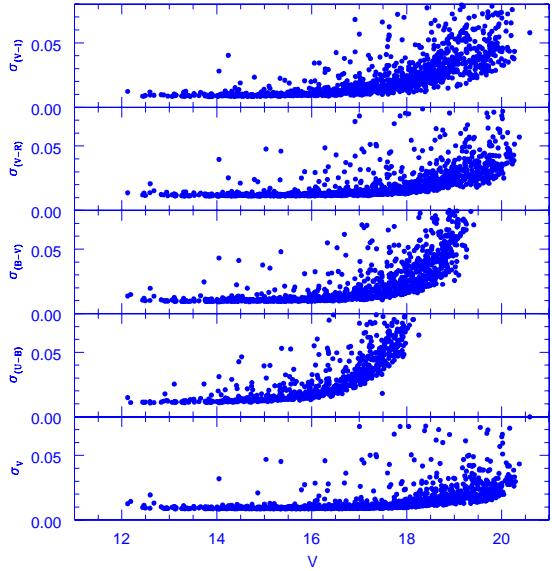,width=9cm,height=9cm}
\caption{Photometric errors in magnitude and colour against $V$ magnitude} 
\end{figure}

\begin{table}
\centering
\caption{Journal of observations, with dates and exposure times for each passband. 
$N$ denotes the number of stars measured in different passband.}
\begin{tabular}{cccc}
\hline
Band  &Exposure Time &Date&$N$\\
&(in seconds)   & &\\
\hline
$U$&1800$\times$2, 300$\times$2&24/25 Feb 2003&770\\
$B$&1200$\times$2, 240$\times$2&,,&1200\\
$V$&900$\times$3, 180$\times$2&,,&1300\\
$R$&500$\times$3, 120$\times$2&,,&1350\\
$I$&300$\times$3, 60$\times$2&,,&1400\\
\hline
\end{tabular}
\end{table}
\subsection{Comparison with previous photometry}
\begin{table}
\centering
\caption{Mean photometric errors in magnitude and colours in different magnitude bins} 
\begin{tabular}{cccccc}
\hline
$V$ &$\sigma_V$   &$\sigma_{U-B}$ &$\sigma_{B-V}$&$\sigma_{V-R}$&$\sigma_{V-I}$\\
&&&&&\\
\hline
12$-$13&0.014&0.012&0.014&0.011&0.010\\
13$-$14&0.010&0.012&0.011&0.012&0.009\\
14$-$15&0.010&0.014&0.011&0.012&0.010\\
15$-$16&0.010&0.017&0.011&0.012&0.011\\
16$-$17&0.010&0.025&0.013&0.013&0.014\\
17$-$18&0.011&0.038&0.017&0.016&0.015\\
18$-$19&0.019&     &0.018&0.025&0.035\\
19$-$20&0.023&     &0.025&0.035&0.054\\
\hline
\end{tabular}
\end{table}

 We have been obtained $UBVRI$ photometry for $\sim$ 1300 stars down to $V=$ 20 mag 
 in the region of NGC 2421. Ramsay et al. (1992) has also presented CCD $UBV$ 
 photometry for 98 stars. Fig 3 shows the 
 plots of difference $\Delta$ in $V$, $(B-V)$ and $(U-B)$ with $V$ magnitude. 
 The average difference (in the sense: our values minus Ramsay et al. (1992)) alongwith 
 their standard deviation are listed in Table 3. The comparison of ours and their 
 $V$ magnitudes shows a weak linear dependence of $\Delta V$ with $V$ which is shown by 
 dotted line in Fig 3. The systematic difference of $\sim -0.03$ mag is present in $(B-V)$ 
 colour without 
 any dependence on stellar magnitude while there is no significant difference or dependency 
 seen in $(U-B)$ colour with stellar magnitude.

\begin{figure}
\centering
\psfig{figure=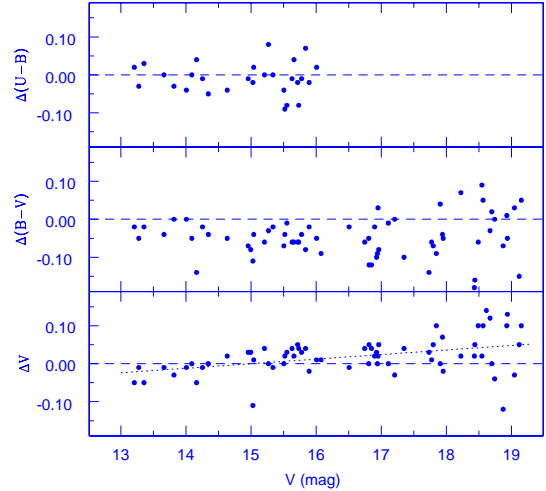,width=9cm,height=9cm}
\caption{Comparison of our $UBV$ photometry with the CCD photometry given by Ramsay et al. (1992). 
Dotted line represents a linear least square fitting to the data points.} 
\end{figure}

\begin{table}
  \caption{Comparison of the present CCD photometry with Ramsay et al. (1992).
  The difference ($\Delta$) is always in the sense present minus
  comparison data. The mean along with their standard deviations in magnitude are based on
  N stars. Few deviated points are not included in the average determination.}
  \begin{tabular}{cccc}
  \hline
  $V$ range&$<\Delta$$V>$&$<\Delta(B-V)$$>$&$<\Delta(U-B)>$\\
  &Mean$\pm$$\sigma$(N)&Mean$\pm$$\sigma$(N)&Mean$\pm$$\sigma$(N)\\
  \hline
  13.0 $-$ 14.0&$-$0.02$\pm$0.02(5)&$-$0.02$\pm$0.02(5)&0.00$\pm$0.02(5)\\
  14.0 $-$ 15.0&0.00$\pm$0.02(8)&$-$0.04$\pm$0.02(8)&0.00$\pm$0.04(8)\\
  15.0 $-$ 16.0&0.02$\pm$0.02(14)&$-$0.04$\pm$0.02(14)&$-$0.01$\pm$0.05(15)\\
  16.0 $-$ 17.0&0.02$\pm$0.02(12)&$-$0.05$\pm$0.04(9)&\\
  17.0 $-$ 18.0&0.02$\pm$0.04(10)&$-$0.04$\pm$0.04(9)&\\
  18.0 $-$ 19.0&0.03$\pm$0.04(8)&$-$0.00$\pm$0.05(10)&\\
  \hline
  \end{tabular}
  \end{table}

\section{Data Analysis}

\subsection{Cluster radius}

 We used radial stellar density profile for the determination of cluster radius. For this, 
 we selected the stars brighter than $V=$ 20.0 mag. The average stellar density was calculated 
 in successive, 50 pixel wide annuli around the cluster center. The cluster center is determined 
 iteratively by calculating average X and Y position of the stars within 400 pixels from an eye 
 estimated center, until they converged to a constant value. In this way, we obtained the pixel 
 coordinate of the cluster center as (425, 400) which is marked by 'C' in the Fig 1. 
 Fig. 4 shows the stellar surface density as a function of distance from the cluster center. The 
 density profile flattens at radius $r=$ 250 pixels. After $r=$ 250 pixels stellar density is 
 merging into the field star density as indicated by horizontal arrow in Fig 4. The surface 
 density of the field stars is derived as 7$\times 10^{-4}$ per pixel$^{2}$. In this way we 
 estimated the radius 3$^\prime$.0 for this cluster which is smaller than the value 
 3$^\prime$.5 given by Mermilliod (1995). 

 We observed 12$^{\prime}$.5$\times$12$^{\prime}$.5 area towards the cluster NGC 2421 which is larger 
 than the cluster radius and hence we have considered the stars as field stars which are having 
 their position more than 1.6 cluster radius (see Fig. 1). The nearest boundary of the field 
 region is about 5$^{\prime}$.0 away from the cluster center in the west direction. 

\begin{figure}
\centering
\psfig{figure=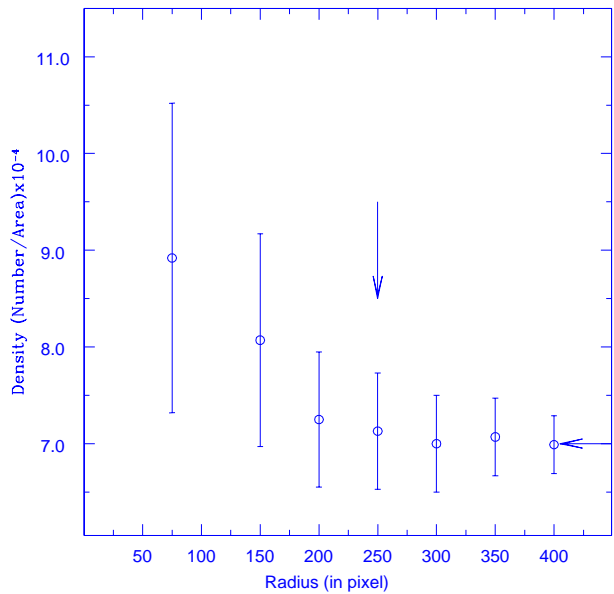,width=10cm,height=10cm}
\caption{Star density as a function of radial distance from the center of the cluster with the 
stars brighter than $V=$ 20.0 mag. Errorbar denotes the error determined 
from sampling statistics(=$\frac{1}{\sqrt{N}}$ where N is the number of stars used in the density
estimation at that point). Horizontal and vertical arrows represent the density of the field 
stars and radius of the cluster respectively.} 
\end{figure}

\subsection{Apparent colour-magnitude diagrams of the cluster and field regions}

 Fig. 5 presents photometric colour-magnitude (CM) diagrams of the cluster and field region. To 
 reduce the field star contamination in the CM diagrams, we used the stars 
 within the cluster radius. The 
 CM diagrams of the cluster extending down to $V \sim$ 19.5 mag except in $V$, $(U-B)$ CM diagram 
 where it is only up to $V \sim$ 18 mag. A main-sequence extending upto $V =$ 19.0 mag is clearly 
 visible in the CM diagrams of the cluster. Contamination due to the field stars 
 is evident, especially in the fainter parts of the CM diagrams. Ideally, one would like to 
 eliminate the 
 contamination from the stars of the galactic field using the information contained in the proper 
 motion/or radial velocity  of the stars. Unfortunately, these studies are not available for this cluster and hence 
 we are forced to base our analysis onto photometric arguments. We selected members by defining 
 the binary sequence. It has been defined by shifting the blue envelope by 0.80 mag vertically, 
 which is shown in the CM diagram of the cluster.
 In Table 4, we have listed the expected number of field stars using $V$, $(V-I)$ CM diagram of the 
 field region. From this Table we can 
 estimate the frequency distribution of stars in different 
 parts of the CM diagram. It is also clear that all photometric probable members can not be cluster 
 members and non-members should be subtracted in the studies of cluster mass function etc. However, 
 probable members located within a cluster radius from its center can be used to determine the 
 cluster parameters, as they have relatively less field star contamination and this has been done 
 in the sections to follow.
 
\begin{figure*}
\centering
\hspace{1.0cm}\psfig{figure=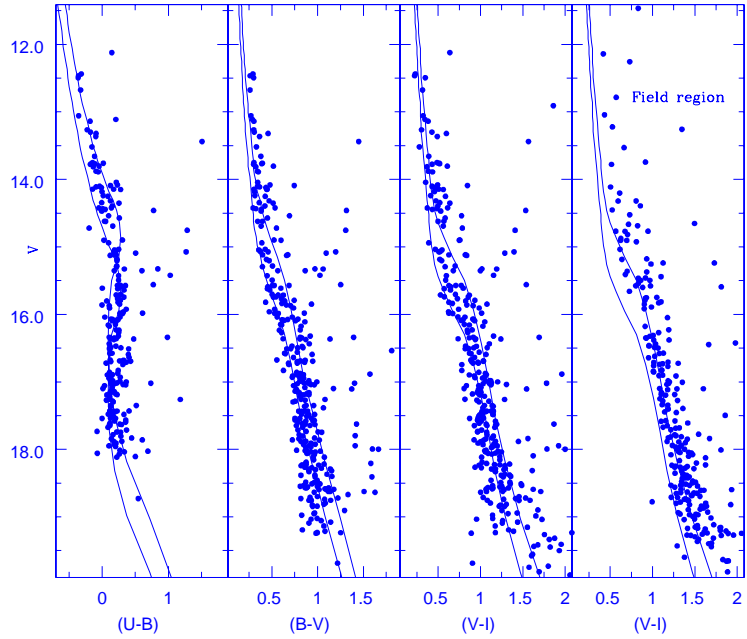,width=10cm,height=10cm}
\caption{ The $V$, $(U-B)$; $V$, $(B-V)$ $V$ and $V$, $(V-I)$ CM 
diagrams for the stars within the cluster radius and $V$, $(V-I)$ CM diagram for the field region. 
Solid lines represent the blue and red envelope of the cluster MS. The red envelope is determined 
by shifting the blue envelope vertically by 0.80 mag.} 
\end{figure*}

\begin{table}
\centering
\caption{Frequency distribution of the stars in the $V$, $(V-I)$
diagram of the cluster and field regions. $N_{B}$, $N_{S}$ and $N_{R}$ denote
the number of stars in a magnitude bin blueward, along and redward of the
cluster sequence respectively. The number of stars in the field regions are 
corrected for area differences. $N_{C}$ (difference between the $N_{S}$ value
of cluster and field regions) denotes the statistically expected number of
cluster members in the corresponding magnitude bin.}

\begin{tabular}{c|ccc|ccc|c}
\hline
&&&&&&&\\
$V$ range &\multicolumn{3}{|c|}{Cluster region} & \multicolumn{3}{|c|}{Field region}& \\
\cline{2-7}
&$N_{B}$&$N_{S}$&$N_{R}$&$N_{B}$&$N_{S}$&$N_{R}$&$N_{C}$ \\
\hline
12 - 13 & 0 & 4& 2 & 0 & 0 &  3& 4 \\
13 - 14 & 0 &13& 7 & 0 & 0 &  6& 13  \\
14 - 15 & 0 &26&10 & 0 & 8 & 10& 18  \\
15 - 16 & 0 &38&14 & 0 & 10& 19& 28  \\
16 - 17 & 1 &52&10 & 0 & 22& 15& 30  \\
17 - 18 & 21&57&11 & 0 & 24& 24& 33 \\
18 - 19 & 24&47&11 & 1 & 37& 35& 10\\
\hline
\end{tabular}
\end{table}

\subsection{Colour-colour diagram}

We present colour-colour (CC) diagram in Fig. 6. The ZAMS given by Schmidt-Kaler (1982) is shown 
by the continuous curve. It is clearly seen that this ZAMS is not fitting to the stars of A and 
F spectral type. Prominent excess in $(U-B)$ colour is clearly visible for the stars of $(B-V) >$ 
0.50 mag. This indicates that the cluster is metal deficient. The UV excess $\delta(U-B)$ 
determined with respect to Hyades MS turns out to be $\sim$ 0.15 mag. We estimated [Fe/H] 
$\sim -$ 0.45 ($Z$ $\sim$ 0.004) adopting [Fe/H] versus $\delta (U-B)$ relation by Carney 
(1979). Furthermore, we fitted the ZAMS given by Bertelli et al. (1994) for $Z$ $=$ 0.004 to determine 
the reddening in the direction of cluster. The ZAMS of $Z$ $=$ 0.004 which is shown by short 
dash lines in the two colour diagram fits well and provides the reddening $E(B-V) = 0.42\pm0.05$ 
for this cluster. Our reddening estimate is in agreement with the earlier findings (see Sec. 1).

 The nature of extinction law has also been studied by using the stars having spectral type 
 earlier than A0. We selected these stars using CC and CM diagram and find that the stars 
 with $V \le$ 15.0 mag and $(B-V) \le$ 0.50 mag are the desired stars. In the absence of 
 spectral class information, we determined their intrinsic colour using $UBV$ photometric 
 Q-method (cf. Johnson \& Morgan (1953)) and the calibrations provided by Caldwell et al. (1993) 
 for $(U-B)_0$, $(V-R)_0$ and $(V-I)_0$ with $(B-V)_0$. Table 5 lists the mean values of the 
 colour excess ratios. These values indicate that law of interstellar extinction is normal in 
 the direction of the cluster.

\begin{table}
\caption{A comparison of the colour excess ratios with $E(B-V)$ with the normal interstellar 
extinction law given by Cardelli et al. (1989).}

\centering
\begin{tabular}{cccc}
\hline
&&&\\
Object&$\frac{E(U-B)}{E(B-V)}$&$\frac{E(V-R)}{E(B-V)}$&$\frac{E(V-I)}{E(B-V)}$\\
&&&\\
\hline
Normal interstellar&0.72&0.65&1.25\\
NGC 2421&0.73$\pm$0.06&0.54$\pm$0.08&1.25$\pm$0.11\\
\hline
\end{tabular}
\end{table}

\begin{figure}
\centering
\psfig{figure=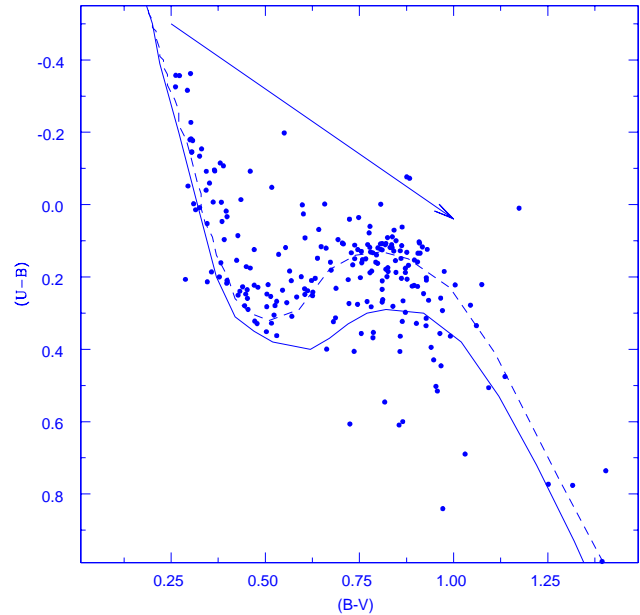,width=9cm,height=9cm}
\caption{ The $(U-B)$ versus $(B-V)$ colour-colour diagram of the cluster. The continuous 
straight line represents slope 0.72 and the direction of reddening vector while continuous 
curve represents locus of Schmidt-Kaler's (1982) ZAMS for solar metallicity. The curve shown 
by short dashed lines is the ZAMS given by Bertelli et al. (1994) for $Z =$ 0.004} 
\end{figure}

\subsection{Interstellar extinction in near-IR}

Two Micron All Sky Survey (2MASS) $JHK_s$ data is available for 180 stars in this cluster and has been used 
for determining the interstellar extinction in the direction of the cluster in near-IR. The data 
is complete up to 16.0 mag in $J$, 15.5 mag $H$ and 15.0 mag in $K_s$. The $K_s$ magnitude 
are converted into $K$ magnitude following Persson et al. (1998). By combining optical and 
near-IR data, we plotted $(J-K)$ versus $(V-K)$ diagram in Fig 7. The ZAMS shown by solid line 
is taken from Bertelli et al. (1994) for $Z =$ 0.004. The fit of ZAMS provides $E(J-K) = 
0.20\pm0.20$ mag
and $E(V-K) = 1.15\pm0.20$ mag for the cluster. The ratio $\frac{E(J-K)}{E(V-K)} \sim 0.18\pm0.30$ 
is in good agreement with the normal interstellar extinction value 0.19 suggested by Cardelli 
et al. (1989).

\begin{figure}
\centering
\psfig{figure=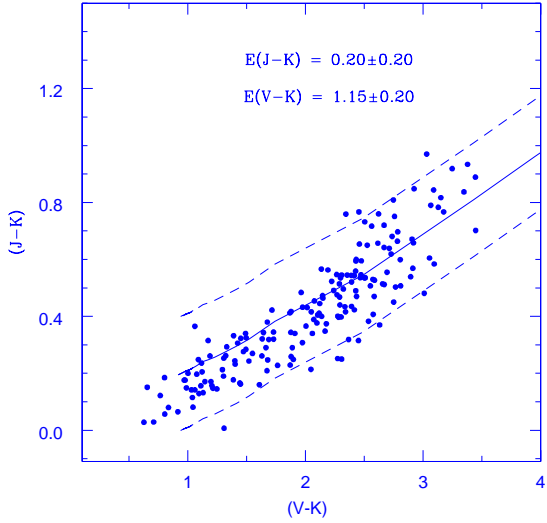,width=10cm,height=10cm}
\caption{ The plot of $(J-K)$ versus $(V-K)$ colour-colour diagram of the cluster for the 
stars within the cluster radius. The solid line is the ZAMS of $Z =$ 0.004 fitted for the 
marked values of the excesses while short dash lines show the errorbars. } 
\end{figure}

\subsection{Colour excess diagram}

In Fig 8, we plotted colour-excess diagram for study of interstellar extinction law using 
optical and near-IR data. Colour excesses have been determined by comparing the observed 
colours of the stars earlier than A0 spectral type (see Sec. 3.3) with its intrinsic colours 
derived from the colour relation given by 
FitzGerald (1970) for $(U-B)$ and $(B-V)$; by Johnson (1966) for $(V-R)$ and $(V-I)$ and 
by Koornneef (1983) for $(V-J)$, $(V-H)$ and $(V-K)$. The colour excesses $E(U-B)$, $E(B-V)$, 
$E(V-R)$, $E(V-I)$, $E(V-H)$, and $E(V-K)$ are plotted against $E(V-J)$ in Fig 8. The solid 
straight line shown in this figure is the least square linear fits to the data point. In all 
the colour-excess diagrams except $E(U-B)$ vs $E(V-J)$ and $E(B-V)$ vs $E(V-J)$, the 
values of correlation coefficient (r) and fit indicate that the data points are well 
represented by linear relation. In $E(U-B)$ vs $E(V-J)$ and $E(B-V)$ vs $E(V-J)$ colour 
excess diagrams, scattering is more pronounced in the colour excess $E(U-B)$ and $E(B-V)$.
In Table 6, we have listed the slope of these straight lines which represent 
the reddening directions in the form of colour excess ratios. The colour excess ratios given 
by Cardelli et al. (1989) for normal interstellar matter are also listed in this Table. 
The present reddening directions generally agree within 3$\sigma$ with those given for the 
normal interstellar extinction law. This indicates that the interstellar extinction law is 
normal in the direction of the cluster. 

\begin{figure}
\centering
\psfig{figure=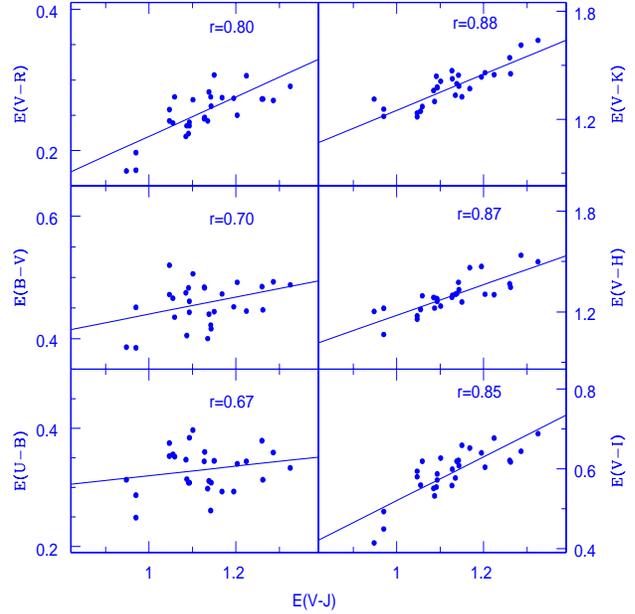,width=9cm,height=10cm}
\caption{ The plot of $E(U-B)$, $E(B-V)$, $E(V-R)$, $E(V-I)$, $E(V-H)$, $E(V-K)$ versus $E(B-V)$,
colour excess diagram of the cluster. The solid straight lines represent the least square linear 
fit to the data point while r denotes its correlation coefficients.}  
\end{figure}

 Furthermore, to know about the nature of interstellar extinction law in the direction of the
 cluster, we have determined the value of R. We used the relation R $=$ 1.1$E(V-K)$/$E(B-V)$ 
 given by Whittet \& Breda (1980) which is generally used for the longer wavelengths. In this 
 way we determined the value of R $= 3.2\pm0.3$ which is in agreement with the value 3.1 for 
 normal extinction law.

 On the basis of above analysis, we can conclude that interstellar extinction law is normal 
 in the direction of the cluster and is in agreement with our earlier result.

\begin{table*}
\caption{A comparison of extinction law in the direction of cluster with normal
extinction law given by Cardelli et al. (1989).}
\begin{center}
\begin{tabular}{cccccccc}
\hline
&&&&&&&\\
Objects&$\frac{E(U-B)}{E(V-J)}$&$\frac{E(B-V)}{E(V-J)}$&$\frac{E(V-R)}{E(V-J)}$&$
\frac{E(V-I)}{E(V-J)}$&$\frac{E(V-H)}{E(V-J)}$&$\frac{E(V-K)}{E(V-J)}$&$\frac{E(
J-K)}{E(V-K)}$\\
&&&&&&&\\
\hline
Normal value&0.32&0.43&0.27&0.56&1.13&1.21&0.19\\
NGC 2421&0.10$\pm$0.09&0.14$\pm$0.10&0.28$\pm$0.05&0.55$\pm$0.08&0.98
$\pm$0.13&1.08$\pm$0.12&0.18$\pm$0.30\\
\hline
\end{tabular}
\end{center}
\end{table*}

\subsection{Near-IR excess fluxes}

 To study the near-IR flux in the stars, we plot $\Delta (V-H)$ and $\Delta (V-K)$ versus 
 $E(V-J)$ in Fig. 9. The value of $\Delta (V-H)$ and $\Delta (V-K)$ has been calculated by 
 taking difference between the observed colour excess in $(V-H)$ and $(V-K)$ based on 
 spectral type and the derived colour excess from $E(V-J)$ assuming normal interstellar 
 extinction law. The possible source of errors may be the observational uncertainties 
 in $JHK$ magnitudes, inaccuracies in the estimation of $E(V-J)$ and in its ratio with 
 $E(V-H)$ and $E(V-K)$; and errors in the spectral and luminosity classifications. 
 Consequently, the differences can be considered statistically significant only if their 
 absolute values are larger than $\sim$ 0.5 mag. Fig. 9 gives an indication that for all 
 the stars the values of $\Delta (V-H)$ and $\Delta (V-K)$ are close to zero. Hence, we 
 can conclude that near-IR fluxes are not seen in any of the star under study indicating 
 absence of gas and dust envelope around them.

\begin{figure}
\centering
\psfig{figure=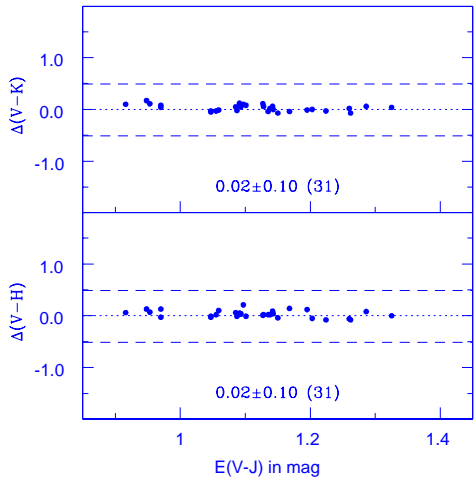,width=10cm,height=10cm}
\caption{ The plot of $\Delta (V-H)$ and $\Delta (V-K)$ versus $E(V-J)$. The horizontal 
dotted lines denote zero excess while short dashed lines denote the extent of the expected 
errors.}
\end{figure}

\subsection{Distance to the cluster}

The distance of the cluster is determined by fitting the ZAMS. The intrinsic CM diagram of the 
cluster is depicted in Fig. 10. In order to reduce the field star contamination, we have used 
only those probable cluster members which are within the cluster radius and photometric 
members (see Sec. 3.2). For converting apparent $V$ magnitude and $(U-B)$,
$(B-V)$, $(V-R)$ and $(V-I)$ colours into intrinsic one, we used average values of
$E(B-V)$ and following relations for $E(U-B)$ (cf. Kamp 1974),
$A_V$ and $E(V-I)$ (Walker 1987) and $E(V-R)$ (Alcal\'{a} et al. 1988).\\

$E(U-B)$ = [X + 0.05$E(B-V)$]$E(B-V)$

\vspace{0.3cm}
where X = 0.62 $-$ 0.3$(B-V)$$_{0}$ for $(B-V)$$_{0}$ $<$ $-$0.09

~~and~~~X = 0.66 + 0.08$(B-V)$$_{0}$ for$(B-V)$$_{0}$ $>$ $-0.09$\\

          $A_V$ = [3.06 + 0.25$(B-V)$$_{0}$ + 0.05$E(B-V)$]$E(B-V)$;\\
	  and      $E(V-R)$ = [E1 + E2E$(B-V)$]$E(B-V)$\\

	  where E1 = 0.6316 + 0.0713$(B-V)$$_{0}$\\
	  and E2 = 0.0362 + 0.0078$(B-V)$$_{0}$;\\

	  $E(V-I)$ = 1.25[1 + 0.06$(B-V)$$_{0}$ + 0.014$E(B-V)$]$E(B-V)$\\

  The ZAMS taken from Bertelli et al. (1994) for $Z =$ 0.004 is plotted in $V_{0}$, $(U-B)_0$; 
  $V_0$, $(B-V)_0$; $V_0$, $(V-R)_0$ 
  and $V_0$, $(V-I)_0$ diagrams. The visual fit of the ZAMS to the 
  bluest envelope of intrinsic CM diagrams provides $(m-M)_0 =$ 11.7$\pm$0.2 mag for the cluster NGC 
  2421. The distance value of the cluster should be reliable since it has been derived by 
  fitting the ZAMS over a wide range of MS. The distance modulus determined above gives a 
  distance of 2.2$\pm$0.2 Kpc for the cluster. Our derived value of the distance is not very 
  much different from the value 1.9 Kpc derived by Moffat \& Vogt (1975) while it is less than 
  the value 2.8 Kpc derived by Ramsay \& Pollacco (1992).
\begin{figure*}
\centering
\hspace{1.0cm}\psfig{figure=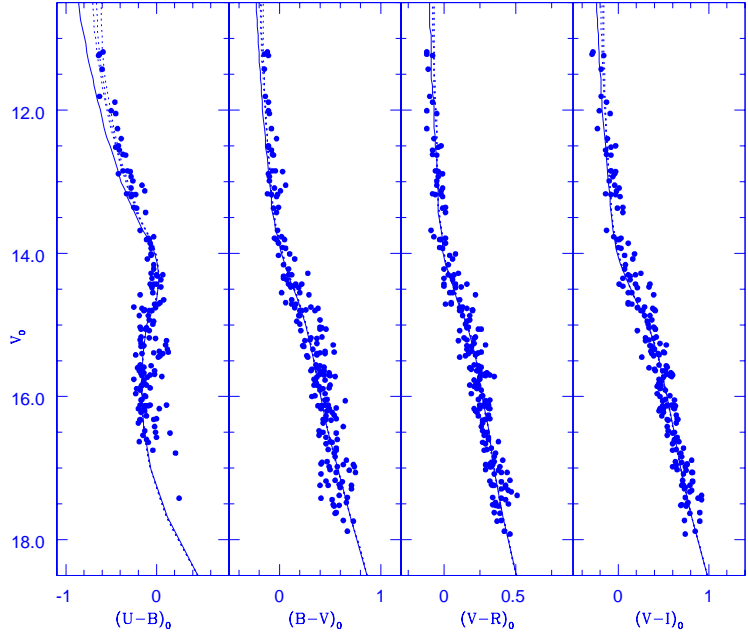,width=10cm,height=10cm}
\caption{ The intrinsic colour-magnitude diagram of the cluster. The continuous solid line 
curve are the ZAMS given by Bertelli et al. (1994) for $Z =$ 0.004. The short dashed line curve 
are the isochrones taken from Bertelli et al. (1994) for $Z =$ 0.004 and log(age) $=$ 7.8,7.9 and 8.0.}
\end{figure*}

\subsection{Age of the cluster}

The age of the cluster is determined by comparing the theoretical stellar evolutionary 
isochrones given by Bertelli et al. (1994) for $Z =$ 0.004 with its intrinsic CM diagram 
(Fig. 10). The isochrones include the effect of mass loss and convective core 
overshooting in the model. We have fitted the isochrones of log(age) $=$ 7.8, 7.9 and 8.0 
to the intrinsic CM diagrams. The isochrones fitted to the brighter stars indicate that 
age of the cluster is 80$\pm$20 Myr.

 To derive the age and distance of the cluster with the combination of optical and near-IR data, 
 we plot $V$ versus $(V-K)$ and $K$ versus $(J-K)$ CM diagram in Fig 11. The 
 theoretical isochrones given by Bertelli et al. (1994) for $Z =$ 0.004 and log(age) $=$ 8.0 have been 
 overplotted in the CM diagram. The apparent distance moduli $(m - M)_{V, (V-K)}$ and
 $(m-M)_{K, (J-K)}$ turn out to be 13.0$\pm$0.3  and 11.8$\pm$0.3 mag for this cluster.
 Using the reddening values estimated in Sec. 3.4, we derived a distance of 2.3$\pm$0.3.
 Both age and distance determination for the cluster are thus in agreement 
 with our earlier estimates. However, scattering is larger due to the large errors in $JHK$ mags.
  
\subsection{Luminosity and Mass function}

To determine the luminosity function of the cluster, we considered $V$ versus $(V-I)$ CM 
diagram in comparison to others because it is deepest. For removing the field star 
contamination, we adopted the photometric criteria by defining the blue and red envelope 
for the MS (see Sec. 3.2). The same envelope is also drawn for the $V$ versus $(V-I)$ 
CM diagram of the field region. Stars are counted within this envelope, for both the 
cluster and the field region CM diagrams. The observed cluster luminosity 
function is the difference between the counts in the two fields after accounting for 
the difference in area between the cluster and field regions.

\begin{figure}
\centering
\psfig{figure=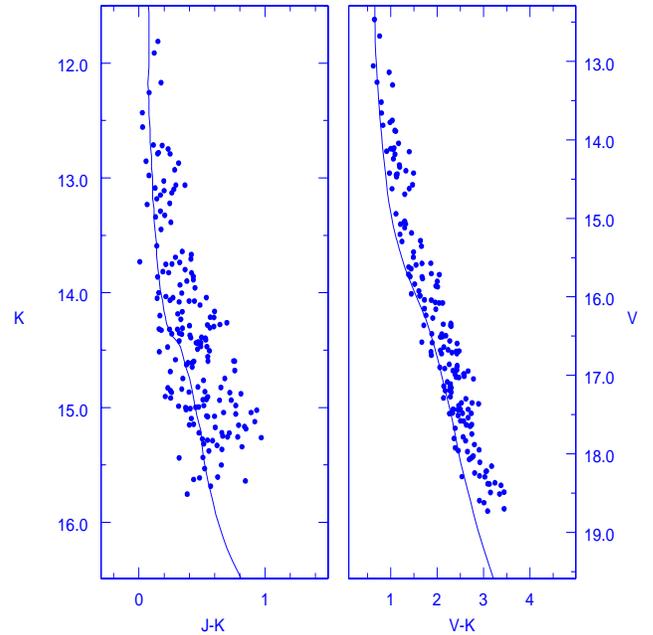,width=9cm,height=10cm}
\caption{ The $K$ versus $(J-K)$ and $V$ versus $(V-K)$ CM diagrams of the cluster using 
probable cluster members. The solid curve represent the isochrones of log(age) $=$ 8.0 taken 
from Bertelli et al. (1994) for $Z =$ 0.004.}
\end{figure}

\begin{table}
\centering
\caption {Variation of completeness factor (CF) in the $V$, $(V-I)$ diagram with
the MS brightness.}
\begin{tabular}{|c|c|}
\hline
$V$ mag range&CF\\
\hline
12 - 13&0.99\\
13 - 14&0.99\\
14 - 15&0.99\\
15 - 16&0.96\\
16 - 17&0.94\\
17 - 18&0.93\\
18 - 19&0.93\\
\hline
\end{tabular}
\end{table}

\begin{figure}
\centering
\psfig{figure=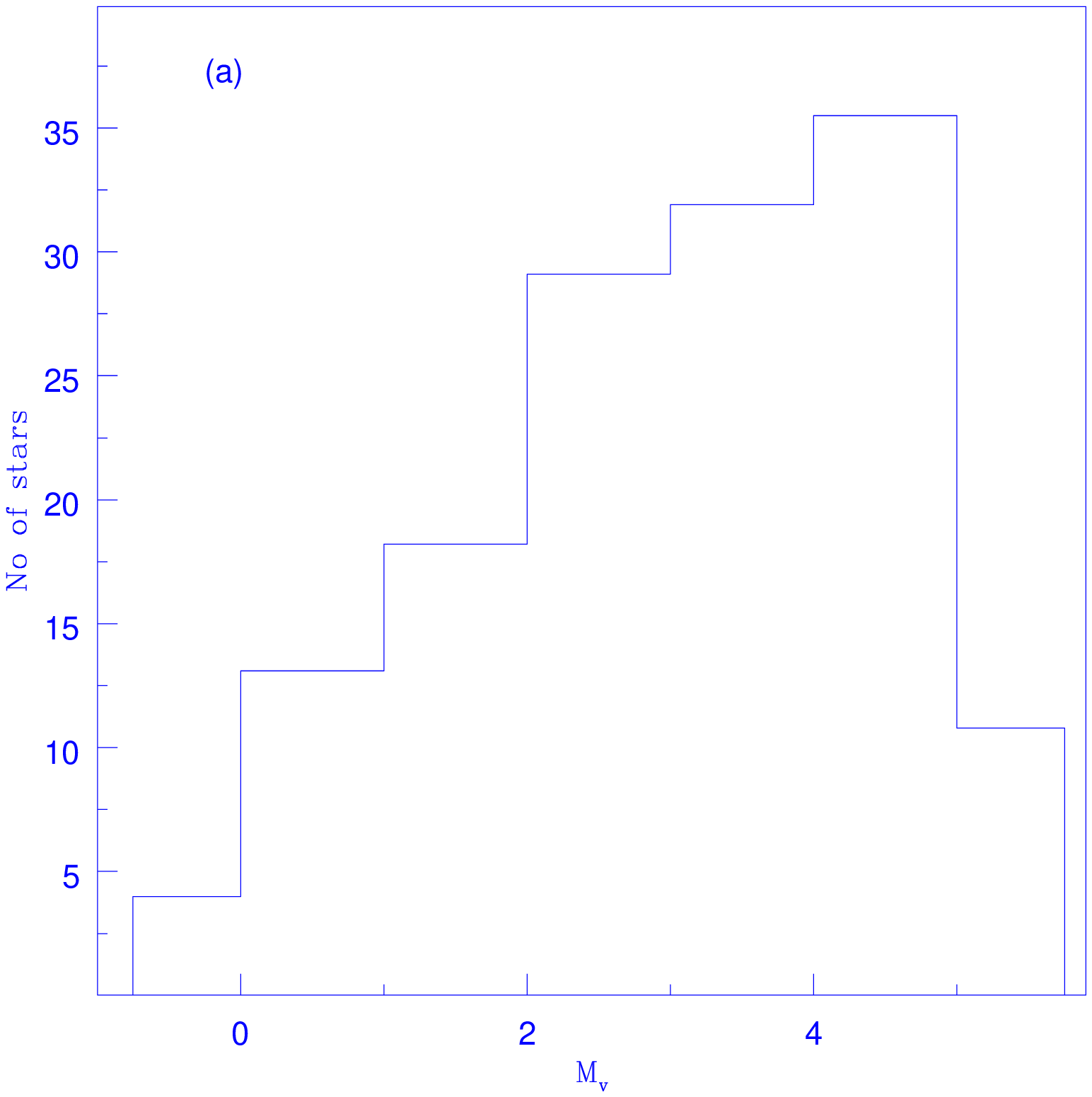,width=9cm,height=9cm}
\psfig{figure=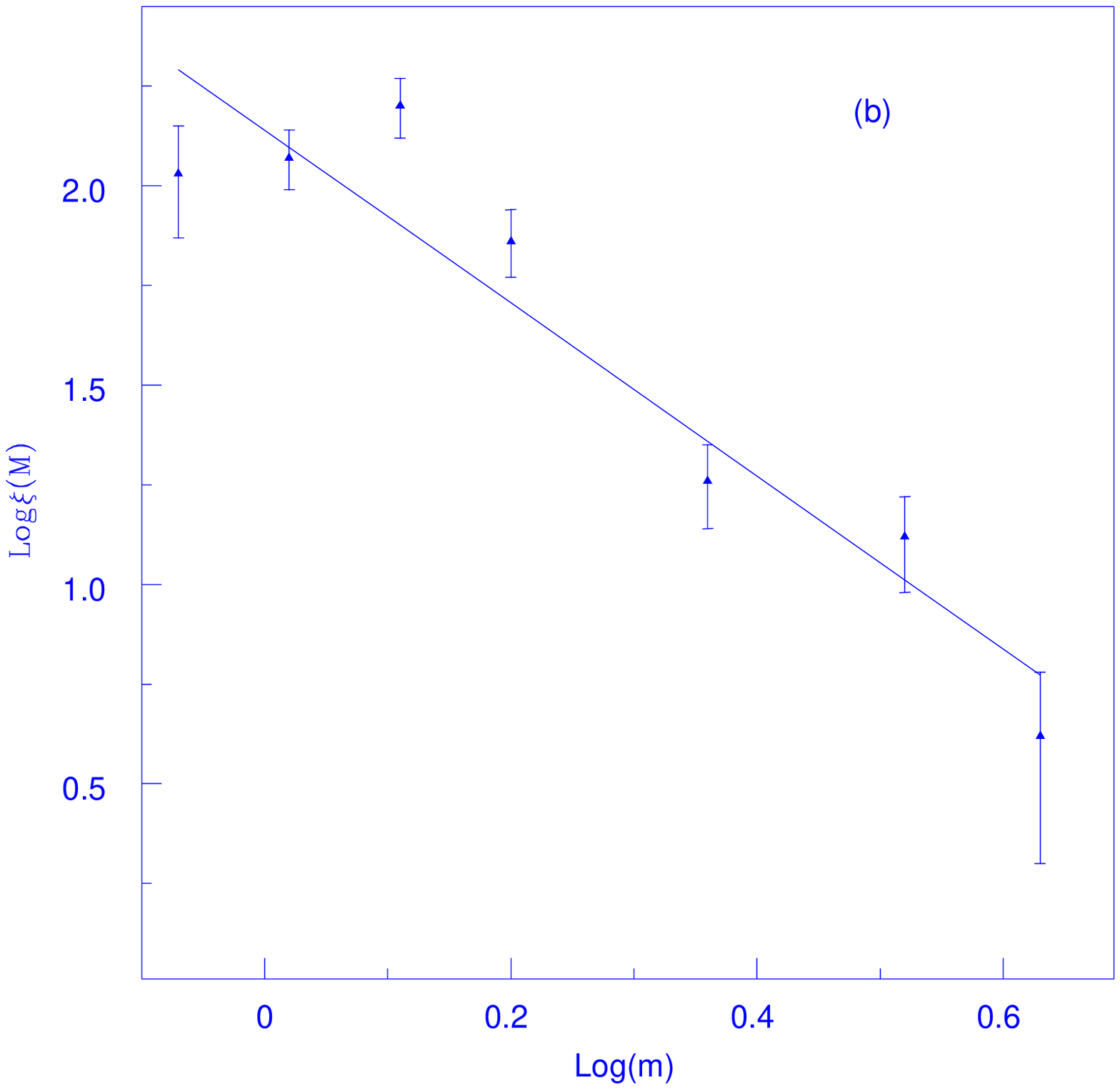,width=9cm,height=9cm}
\caption{ (a) Luminosity function of the cluster. (b) Mass function derived using Bertelli 
et al. (1994) isochrones.}
\end{figure}

 Incompleteness corrections are handled by introducing a number of artificial stars into 
 the original data images of $V$ and $I$ passband. The method used for this has been 
 described in Yadav \& Sagar (2002), so we present only the results here. Table 7 shows 
 the completeness factor in the cluster region. To summarize, the completeness in the 
 cluster region for MS stars is found to be 93.0\% at $V =$ 19.0 mag. The completeness in 
 the field region has been considered as 100\%.

 The final corrected star counts are found by applying the incompleteness corrections in 
 the cluster region. Fig. 12(a) shows the final luminosity function for the cluster NGC 2421.
 The luminosity function of the cluster rises until $M_v =$ 4.5 and then decreases. 

 The mass function (MF) can be derived by using the relation 
 log$\frac{dN}{dM}$ = $-$(1+$x$)$\times$log($M$)$+$constant, where $dN$ represents the 
 number of stars in a mass bin $dM$ with central $M$ and $x$ is the slope of MF. In 
 transfering the LF to the MF, we need theoretical evolutionary tracks and accurate 
 knowledge of cluster parameters like reddening, distance, age, etc. Theoretical models 
 given by Bertelli et al. (1994) has been used to convert LF to MF and the resulting MF is shown in Fig. 12(b).
  The derived slope of the MF is $x$ = 1.2$\pm$0.3 is in agreement within the error with 
  the value 1.35 given by Salpeter (1955) for the Solar neighbourhood stars.

\subsection{Dynamical state and Mass segregation}

 In the lifetime of a star cluster, encounters between its member stars gradually lead to an 
 increased degree of energy equipartition throughout the cluster. The most significant 
 consequence of this process is that the higher-mass cluster stars gradually sink towards the 
 cluster center and in the process transfer their kinetic energy to the more numerous lower-mass 
 stellar component, thus leading to mass segregation. The time-scale on which a cluster will 
 have lost all traces of its initial conditions is well represented by relaxation time $T_E$. 
 It is given by

\begin{center}
\begin{displaymath}
\hspace{2.0cm}T_{E} = \frac {8.9 \times 10^{5} N^{1/2} R_{h}^{3/2}}{ <m>^{1/2}log(0.4N)}
\end{displaymath}
\end{center}

 where $N$ is the number of cluster members, $R$$_{h}$ is the half-mass radius
 of the cluster and $<m>$ is the mean mass of the cluster stars (cf.
 Spitzer \& Hart 1971). The value of $R$$_{h}$ has been assumed as half of the cluster radius 
 derived by us. Using distance, the angular value of the radius has been converted 
 into linear value. The probable cluster members are selected using CM diagram of the cluster 
 after removing the field star contamination and applying data incompleteness corrections. In 
 this way, we estimated the dynamical relaxation time $T_E =$ 30 Myr for this cluster which 
 implies that the cluster age is 2.6 times its relaxation age. Therefore, we can conclude that 
 the cluster is dynamically relaxed.

  To investigate sign of mass segregation, we split the cluster members in three mass 
  range 4.7$\le$ M$_{\odot}$$<$3.4, 3.4$\le$ M$_{\odot}$$<$2.1 and 2.1$\le$ M$_{\odot}$$\le$0.8. 
  Fig. 13 
  shows the cumulative radial stellar distribution of stars for different masses. To examine 
  the distribution of stars whether they belong to the same distribution or not, we performed 
  K-S test among these distribution. The K-S test provides sign of mass segregation at 
  confidence level of 85\%. In the previous paragraph, we have seen that this cluster is 
  dynamically relaxed and hence one of the possible cause of mass segregation may be the 
  dynamical evolution of the cluster.

\begin{figure}
\centering
\psfig{figure=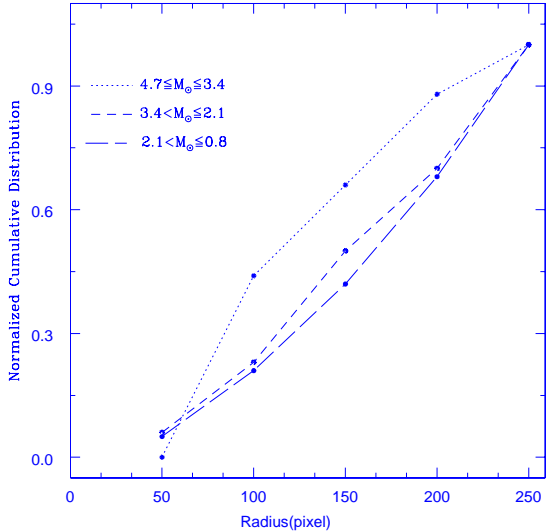,width=9cm,height=9cm}
\caption{ The cumulative radial distribution of stars in various mass range.}
\end{figure}

\section{Conclusions}

 We have investigated the area of open cluster NGC 2421 using $UBVRI$ CCD and 2MASS 
 $JHK_s$ data. The main results of our analysis are the following.

   \begin{enumerate}
      \item The radius of the cluster is determined as 3$^\prime$.0 which corresponds 
      to 1.9 pc at the distance of the cluster.
      \item We estimated the abundance of the cluster stars as $Z =$ 0.004 using excess in 
      $(U-B)$. The $(U-B)$ versus $(B-V)$ colour-colour diagram yields $E(B-V) = 0.42\pm$0.05.
      The analysis of $JHK$ data in combination with the optical data provide $E(J-K) = 
      0.20\pm$0.20 mag
      and $E(V-K) = 1.15\pm$0.20 mag. Our analysis shows that interstellar extinction law is 
      normal towards the cluster direction. No stars found which are having near-IR fluxes 
      due to the presence of circumstellar material around them.
      \item A ZAMS fitting procedure gives a distance of 2.2$\pm$0.2 Kpc for this cluster 
      which is also supported by the value of 2.3$\pm$0.3 Kpc determined by us using the optical and 
      near-IR data. An age of 80$\pm$20 Myr is determined using the isochrones of $Z =$ 0.004
      given by Bertelli et al. (1994).
      \item The mass function slope $x = 1.2\pm0.3$ is derived by considering the corrections 
      of field star contamination and data incompleteness. Our analysis indicate that the cluster
       NGC 2421 is dynamically relaxed and one plausible reason of this relaxation may be the 
       dynamical evolution of the cluster.
   \end{enumerate}

\section*{Acknowledgements}
We thank the referee for valuable comments, which have improved the quality of this paper. This 
study made use of 2MASS and WEBDA.

\end{document}